\documentclass[aps,prd,reprint,superscriptaddress]{revtex4-2}

\usepackage{graphicx}
\usepackage{orcidlink}
\usepackage{amsmath}

\begin{document}
\title{SGNL: Scalable Low-Latency Gravitational Wave Detection Pipeline for Compact Binary Mergers}
\author{Yun-Jing Huang \orcidlink{0000-0002-2952-8429}}
\email{yun-jing.huang@ligo.org}
\affiliation{Department of Physics, The Pennsylvania State University, University Park, PA 16802, USA}
\affiliation{Institute for Gravitation and the Cosmos, The Pennsylvania State University, University Park, PA 16802, USA}

\author{Chad Hanna \orcidlink{0000-0002-0965-7493}}
\affiliation{Department of Physics, The Pennsylvania State University, University Park, PA 16802, USA}
\affiliation{Institute for Gravitation and the Cosmos, The Pennsylvania State University, University Park, PA 16802, USA}
\affiliation{Department of Astronomy and Astrophysics, The Pennsylvania State University, University Park, PA 16802, USA}
\affiliation{Institute for Computational and Data Sciences, The Pennsylvania State University, University Park, PA 16802, USA}

\author{Leo Tsukada  \orcidlink{0000-0003-0596-5648}}
\affiliation{Department of Physics and Astronomy, University of Nevada, Las Vegas, 4505 South Maryland Parkway, Las Vegas, NV 89154, USA}
\affiliation{Nevada Center for Astrophysics, University of Nevada, Las Vegas, Las Vegas, NV 89154, USA}

\author{Amanda Baylor \orcidlink{0000-0003-0918-0864}}
\affiliation{Leonard E.\ Parker Center for Gravitation, Cosmology, and Astrophysics, University of Wisconsin-Milwaukee, Milwaukee, WI 53201, USA}

\author{Olivia Godwin \orcidlink{0000-0002-7489-4751}}
\affiliation{LIGO Laboratory, California Institute of Technology, MS 100-36, Pasadena, California 91125, USA}

\author{Prathamesh Joshi \orcidlink{0000-0002-4148-4932}}
\affiliation{Department of Physics, The Pennsylvania State University, University Park, PA 16802, USA}
\affiliation{Institute for Gravitation and the Cosmos, The Pennsylvania State University, University Park, PA 16802, USA}
\affiliation{School of Physics, Georgia Institute of Technology, Atlanta, GA 30332, USA}

\author{James Kennington \orcidlink{0000-0002-6899-3833}}
\affiliation{Department of Physics, The Pennsylvania State University, University Park, PA 16802, USA}
\affiliation{Institute for Gravitation and the Cosmos, The Pennsylvania State University, University Park, PA 16802, USA}

\author{Cody Messick \orcidlink{0000-0002-8230-3309}}
\affiliation{Leonard E.\ Parker Center for Gravitation, Cosmology, and Astrophysics, University of Wisconsin-Milwaukee, Milwaukee, WI 53201, USA}

\author{Surabhi Sachdev \orcidlink{0000-0002-0525-2317}}
\affiliation{School of Physics, Georgia Institute of Technology, Atlanta, GA 30332, USA}

\author{Ron Tapia}
\affiliation{Department of Physics, The Pennsylvania State University, University Park, PA 16802, USA}
\affiliation{Institute for Computational and Data Sciences, The Pennsylvania State University, University Park, PA 16802, USA}

\author{Zach Yarbrough \orcidlink{0000-0002-9825-1136}}
\affiliation{Department of Physics and Astronomy, Louisiana State University, Baton Rouge, LA 70803, USA}

\begin{abstract}
    We present SGNL, a scalable, low-latency gravitational-wave search pipeline. It reimplements the core matched-filtering principles of the GstLAL pipeline within a modernized framework. The Stream Graph Navigator library, a lightweight Python streaming framework, replaces GstLAL's GStreamer infrastructure, simplifying pipeline construction and enabling flexible, modular graph design. The filtering core is reimplemented in PyTorch, allowing SGNL to leverage GPU acceleration for improved computational scalability. We describe the pipeline architecture and introduce a novel implementation of the Low-Latency Online Inspiral Detection algorithm in which components are pre-synchronized to reduce latency. Results from 40 days of data show that SGNL's event recovery and sensitivity are consistent with GstLAL's within statistical and systematic uncertainties. Notably, SGNL achieves a median latency of 4.7 seconds, compared to 9.0 seconds for GstLAL.
    
\end{abstract}

\maketitle

\section{Introduction}
The fifth Gravitational-Wave Transient Catalog (GWTC-5.0) \cite{gwtc-5} now includes 390 candidate events observed by the LIGO-Virgo-KAGRA (LVK) Collaboration \cite{ligo,virgo,kagra}, combining discoveries from previous observing runs \cite{gwtc-1,gwtc-2,gwtc-3} and the first \cite{gwtc-4} and second parts of the fourth observing run (O4), and illustrating the continued increase in detection rates. During the O4 run, over 250 significant alerts have already been issued in low latency \cite{gracedb}. As detector sensitivities improve and the global detector network expands \cite{kagra, indigo1, indigo2}, the rate of real-time detections is expected to grow further, highlighting the need for pipelines that are fast, scalable, and sustainable over the long term.

The events reported in gravitational-wave catalogs and public alerts are identified by several transient searches. Modeled compact-binary-coalescence (CBC) searches use banks of waveform templates to identify compact-binary mergers, with pipelines such as GstLAL \cite{cannon2012a,cannon2012b,messick2017,sachdev2019,hanna2020,cannon2021,sakon,ray2023,tsukada,ewing2024,joshi2025a,joshi2025b}, MBTA \cite{mbta1,mbta2,mbta3,mbta4}, PyCBC \cite{pycbc1,pycbc2,pycbc3,pycbc4}, and SPIIR \cite{spiir1,spiir2,spiir3} contributing to LVK observing runs. These pipelines differ in their filtering implementations and background estimation methods. Other searches, such as coherent WaveBurst (cWB) \cite{cwb1,cwb2,cwb3}, look for short-duration excess power without relying on a specific compact-binary waveform model, making them sensitive to a broader class of transient sources. Machine-learning-based searches, including MLy \cite{mly} and Aframe \cite{aframe1,aframe2}, have also been developed for gravitational-wave transient detection.

The GstLAL low-latency search pipeline has been central to LVK observing runs, including performing the first low-latency detection of GW170817 \cite{gw170817}, and it continues to identify low-latency events throughout O4 \cite{gracedb}. GstLAL is a stream-based matched-filtering pipeline \cite{messick2017,cannon2012b} in which data are processed via GStreamer \cite{gstreamer}, a C-based multimedia framework. Its core matched-filtering engine employs the Low-Latency Online Inspiral Detection (LLOID) algorithm \cite{cannon2012b}, which splits templates into downsampled time slices and performs singular value decomposition (SVD) \cite{svd} on groups of templates. The full pipeline, including data handling and LLOID filtering, is constructed by connecting GStreamer elements into a processing graph. While GStreamer is powerful and robust, extending GstLAL requires writing new components in C, which must be compiled, manage memory manually, and include additional setup to register each component within the framework. This makes the pipeline slow to develop and difficult to maintain. Its matched filtering also runs on CPUs, limiting the use of modern computing hardware such as graphics processing units (GPUs).

The SGNL pipeline, short for Stream Graph Navigator inspiraL, implements these ideas within a modern, modular, and sustainable framework. It replaces GstLAL’s GStreamer infrastructure with the Stream Graph Navigator (SGN) library \cite{sgn}, a lightweight Python framework for constructing real-time dataflow graphs. SGN organizes computation into modular elements, enabling flexible pipeline design, precise synchronization, and efficient utilization of CPU and GPU resources. At its core, the matched filtering algorithm in SGNL is implemented in PyTorch, allowing GPU acceleration and batch operations across multiple detectors and template banks, which has been shown to scale effectively on GPUs \cite{huang2025}. This scalability will be valuable in future searches, where it can enable the use of larger template banks to explore a wider parameter space. This combination of a Python-based streaming architecture and GPU-accelerated computation provides a maintainable and extensible platform for low-latency gravitational-wave detection.

In this paper, we present the design and validation of the SGNL pipeline. Section~\ref{sec:pipeline} details the architecture of the SGNL online inspiral analysis, including data reading, whitening, the reimplementation of the LLOID filtering algorithm, and event identification. We introduce a novel pre-synchronization scheme that prevents additional latency during multirate reconstruction. Section~\ref{sec:results} evaluates SGNL’s performance using a 40-day Mock Data Challenge (MDC), comparing its sensitivity, event recovery, and latency against GstLAL. Section~\ref{sec:conclusion} concludes with a summary and outlook for future developments.

\section{Pipeline description}\label{sec:pipeline}
\begin{figure*}
    \centering
    \includegraphics[width=0.75\linewidth]{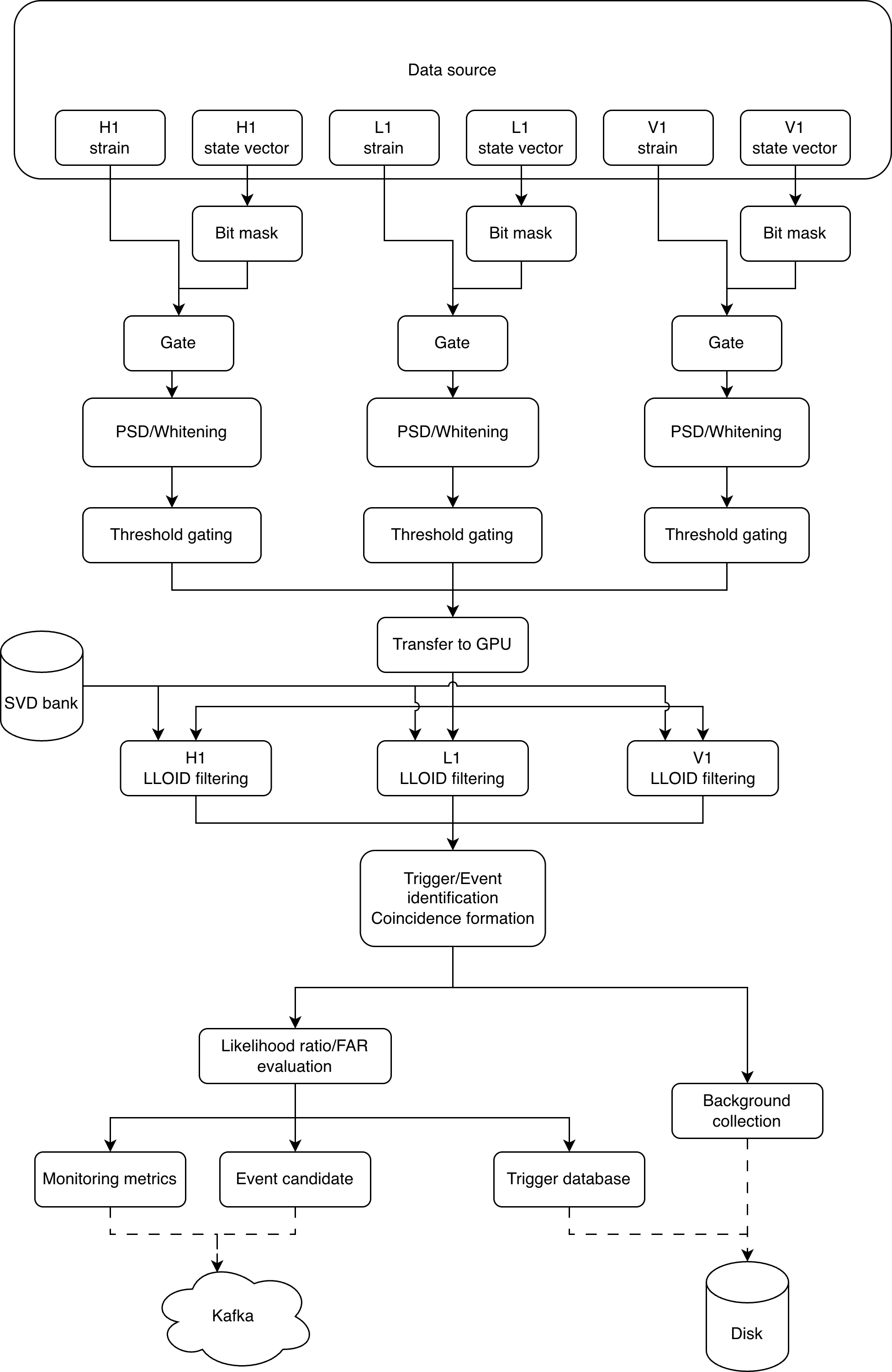}
    \caption{Workflow of the SGNL online inspiral pipeline. Data from multiple detectors are read in a single source element, including both strain and state vector channels. Invalid segments flagged by the state vector are gated from the strain data. The PSD is estimated, and data are whitened. Noise transients above a threshold are gated out. Conditioned data are synchronized across detectors and transferred to the GPU if in GPU mode. Each detector’s data are matched-filtered with a pregenerated SVD bank to produce SNR time series. SNR peaks are identified as triggers and combined into coincident events across detectors. Likelihood ratios and FAR are computed. Non-coincident triggers form background data and are saved to disk. The best events are sent to Kafka for aggregation and GraceDB alerts. Event candidates are saved to an SQLite database on disk.}
    \label{fig:online}
\end{figure*}
In this section, we describe the SGNL online inspiral pipeline and its key components. In Section \ref{sec:sgn}, we discuss the Python streaming framework (SGN) on which SGNL is built. Section \ref{sec:datacond} covers data reading and conditioning, including the handling of external data, power spectral density (PSD) estimation and whitening, and the removal of invalid segments and glitches. In Section \ref{sec:lloid}, we describe the LLOID algorithm and its implementation in SGNL, including a novel pre-synchronization technique to avoid additional latency. Section \ref{sec:event} explains event identification, covering signal-to-noise ratio (SNR) peak detection, coincidence formation, significance estimation, background collection, and output of event candidates. Fig. \ref{fig:online} illustrates the overall workflow of the SGNL online inspiral pipeline. The offline pipeline closely mirrors the online pipeline through filtering and trigger generation, but differs in upstream data access, significance estimation, and the low-latency alerting infrastructure, which are specific to the online pipeline.

\subsection{Stream Graph Navigator}\label{sec:sgn}

The SGNL pipeline is built on the lightweight Python streaming framework SGN. In contrast, the GstLAL pipeline uses GStreamer, a C-based multimedia framework. SGN adapts GStreamer’s core ideas into a simpler and more flexible form for Python. It organizes computation into modular elements connected through source and sink pads. Pipelines form directed acyclic graphs where data flow continuously between stages. This design makes complex workflows easy to define and efficient to run in real time.

The SGN-TS \cite{sgn} extension builds on this framework. It adds explicit notions of time and time series tracking. It also provides tools for buffering data and performing signal processing tasks such as resampling, correlation, synchronization, and stream combination. These features support a wide range of streaming applications. They are especially useful for gravitational-wave analysis, where precise alignment and coincidence of detector signals are crucial.

The SGN-LIGO \cite{sgn-ligo_code} extension builds further on this system. It integrates LVK software such as LALSuite \cite{lalsuite} by wrapping standard tools as SGN elements. This allows gravitational-wave analysis routines to run directly inside the streaming pipeline. It removes the need for separate preprocessing steps or manual data conversions. The result is a simpler and more unified system where all components operate together in real time.

At the top level, SGNL assembles these components into a complete gravitational-wave search pipeline. Each stage, including data reading, conditioning, whitening, matched filtering, and event identification, is implemented as a connected network of SGN elements. Data flow continuously from input to output. This modular design balances flexibility and performance. It also makes the pipeline easy to extend by adding or modifying elements without disrupting other components.

\subsection{Data reading and conditioning}\label{sec:datacond}
\subsubsection{Data reading}\label{sec:data}

Gravitational-wave data from the LVK detectors are distributed in low latency to shared memory on the LIGO Data Grid in one-second chunks as Gravitational-Wave Frame (GWF) files, typically at a sample rate of 16384 Hz. These data contain both gravitational-wave strain and state vector channels, the latter providing data quality information from auxiliary detector channels. The SGNL low-latency pipeline begins with a source element that reads these GWF files and streams both channels to downstream elements in one-second buffers. Unlike GstLAL, which uses a separate source element for each detector, SGNL employs a single unified source element for all detectors. This design allows the source to re-establish synchronization whenever data delivery to shared memory is misaligned across detectors and prevents additional latency when detector streams fall out of sync.

Because low-latency data can experience network or delivery delays, the source element waits up to 60 seconds for missing data before inserting gap buffers to represent missing segments. During this period, it emits heartbeat buffers to maintain activity. After 60 seconds, it begins emitting one-second gap buffers at a one-second cadence to maintain a steady 60-second lag. When new data arrive, the source resumes normal output if the data are continuous. Otherwise, it continues emitting one-second gap buffers as quickly as the pipeline can process them until the discontinuity is fully bridged.

The state vector channel is also used to mask out strain data. This is performed by a gating operation, which gates the strain based on an external control signal. In the online pipeline this control signal is the state vector, where segments flagged as invalid are gated and replaced with gaps, ensuring that only valid data are passed downstream for analysis.

\subsubsection{PSD estimation and whitening}
\begin{figure}
    \centering
    \includegraphics[width=1\linewidth]{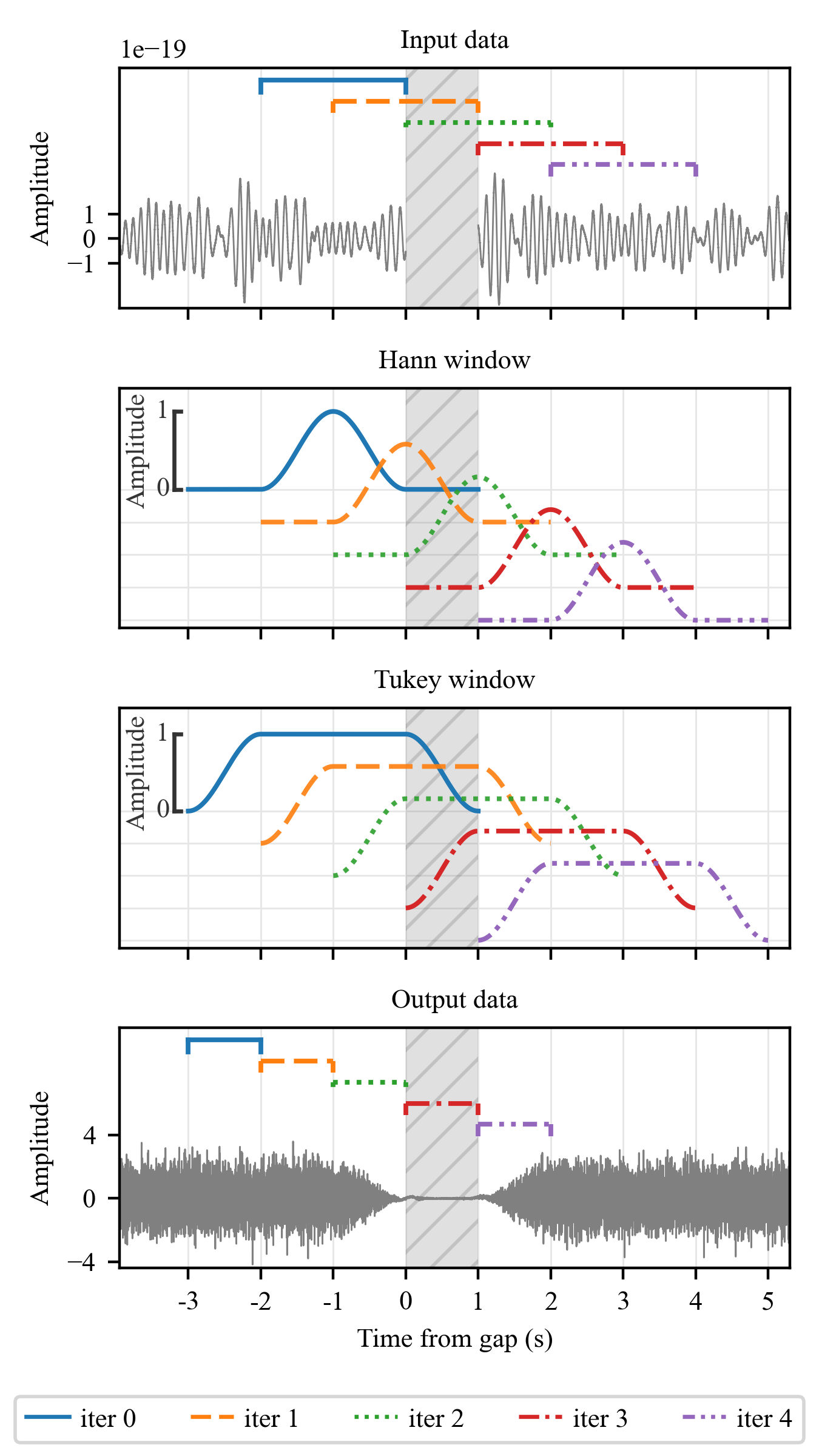}
    \caption{Overlapping and windowing scheme used in the PSD estimation and whitening process, including the behavior around a gap in the input data. In this example, N is set to four seconds of samples, and Z is set to one second. The top panel shows the input stream, a stretch of LIGO Hanford strain from the MDC data described in Sec.~\ref{sec:results}, with a one-second gap, marked by the shaded region between 0 and 1 second, manually inserted to illustrate the gap handling. Brackets indicate five consecutive processing iterations, distinguished by color and line style (legend). Each two-second input block is multiplied by a Hann window of the same line style (second panel), zero-padded to form a four-second FFT block, and transformed to estimate the PSD. The second panel illustrates the overlapping Hann windows, which sum to unity, while the third panel shows the overlapping Tukey windows applied after whitening and inverse FFT. The Hann and Tukey windows for iterations 1 and 2 are shown for illustration only. Because their two-second input segments overlap the gap, no PSD is estimated for those blocks, and the windows are not used in practice. The final panel shows the whitened output of each iteration, each lagging the input by two seconds, corresponding to the two-second latency introduced by the four-second FFT length configuration. Whitened data continue to be produced across the gap as the overlap from earlier iterations drains.}
    \label{fig:hann}
\end{figure}
\begin{figure}
    \centering
    \includegraphics[width=0.8\linewidth]{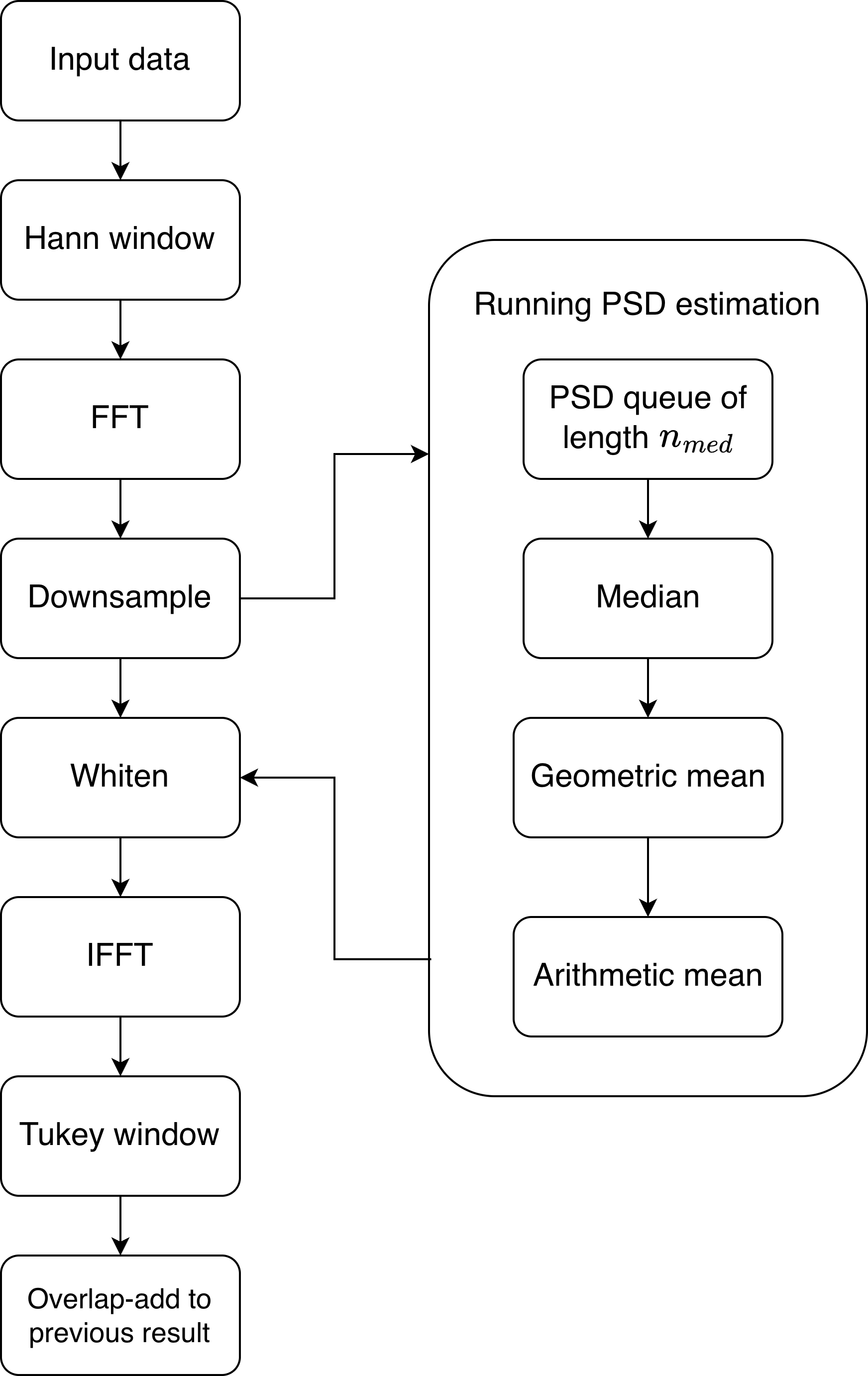}
    \caption{Workflow of the whitening and PSD estimation in SGNL. Input data are first multiplied by a zero-padded Hann window, and an FFT is performed. Unlike GstLAL \cite{tsukada2017}, downsampling is applied after the FFT, where high-frequency components are removed in the frequency domain. The resulting data are used both to generate the whitened output and to update the running PSD estimate. After whitening with the latest PSD, an inverse FFT is applied, followed by a Tukey window. The processed block is then combined with results from previous iterations using overlap-add, and the portion that no longer overlaps is output as the final whitened data.}
    \label{fig:whitening}
\end{figure}
The SGNL pipeline reimplements the GstLAL PSD estimation and whitening method~\cite{messick2017, tsukada2017} in Python within an SGN element. PSD estimation and whitening are performed on fast Fourier transform (FFT) blocks of length $N$, constructed from overlapping input segments. Each input segment of length $N - 2Z$ is multiplied by a Hann window and zero-padded with $Z$ points on each side to form a length-$N$ block for the FFT computation. This reduces spectral leakage and ensures that overlapping windows sum to unity. The windowed blocks are Fourier transformed to produce instantaneous PSD estimates. 
For each frequency bin, the median of the most recent $n_{\rm med}$ PSD estimates is used to update a running geometric mean. Assuming stationary Gaussian noise, the PSD bins are $\chi^2$-distributed, so the geometric mean can be inferred from the median by dividing by a constant factor. To whiten the data, this geometric mean is converted to an arithmetic mean by multiplying by $\exp(\gamma)$, where $\gamma$ is Euler's constant, giving the PSD estimate \cite{messick2017}.

For data whitening, each length-$N$ FFT block is divided by the square root of the PSD to generate the whitened data. An inverse FFT is then applied to obtain the whitened data in the time domain, which is tapered with a Tukey window. Successive blocks are combined using the overlap-add method, with each new Tukey-windowed block added to the previous one with an overlap of $N/2 + Z$. The first $N/2 - Z$ samples of each processed block constitute the final output whitened data, which therefore lag the input by $N/2$ points. The input segment is then advanced by $N/2 - Z$ points for the next iteration.

The FFT block and overlap scheme introduce a latency of N/2 sample points in PSD estimation and whitening. For low-latency operation, N is usually four seconds of samples, Z is one second, and $n_{med} = 7$. This gives an effective latency of two seconds. Fig. \ref{fig:hann} illustrates how the input, Hann window, Tukey window, and output blocks overlap and advance.

Strain data, typically sampled at 16,384 Hz, must be downsampled to match the maximum sample rate of the template bank, usually 2,048 Hz. In SGNL, downsampling is performed in the frequency domain after the FFT by removing high-frequency components. In contrast, GstLAL performs downsampling before the strain data enter the PSD estimation and whitening stage, using a finite impulse response (FIR) filter (see Sec.~\ref{sec:down}) in the time domain. This FIR-based approach adds extra latency, whereas frequency-domain downsampling during the FFT avoids it. Fig. \ref{fig:whitening} illustrates the full PSD estimation and whitening workflow.

Fig. \ref{fig:hann} also shows the effect of data gaps. When gaps occur, any FFT block with input data containing gaps is skipped, and no PSD is estimated for that segment (iterations 1 and 2 in Fig. \ref{fig:hann}). However, whitened data continue to be produced until the overlap from previous iterations is fully drained.

\subsubsection{Threshold gating}
LVK detector strain data contain non-Gaussian, short-duration noise transients, or ``glitches," which can mimic gravitational-wave signals from high-mass compact binaries. The SGNL pipeline employs the same gating method as GstLAL to address these artifacts \cite{messick2017, ewing2024}. Once the data are whitened, giving them unit variance, any brief excursion exceeding a threshold defined in standard deviations ($\sigma$) is set to a gap buffer, with a 0.5s padding applied on either side.

\subsubsection{Transferring data to GPU}
In SGNL’s GPU mode, whitened and gated data from each detector are synchronized and transferred to the GPU. If half-precision computation is enabled, the data are converted to half-precision at this stage.
When running in CPU mode, this step is skipped and the data pass through unchanged.

\subsection{The LLOID filtering algorithm}\label{sec:lloid}

The SGNL analysis builds upon the time-domain matched filtering method used in GstLAL, in which compact binary coalescence signals are identified by correlating the detector strain time series with a large bank of template waveforms \cite{messick2017, cannon2012b}. In the time domain, the matched filter output for a template $h_i(t)$ and detector data $d(t)$ is given by
\begin{equation}\label{eq:mf}
    x_i(t) = 2\int_{-\infty}^{\infty} \hat{h}_i(\tau)\, \hat{d}(t + \tau)\, d\tau,i\in[0, 2M-1],
\end{equation}
where $M$ is the number of templates, the whitened data
\begin{equation}
    \hat{d}(\tau) = \int_{-\infty}^{\infty} df\, \frac{\tilde{d}(f)}{\sqrt{S_n(|f|)}}\, e^{2\pi i f \tau},
\end{equation}
and the whitened templates $\hat{h}_i(\tau)$ are defined similarly. The LLOID algorithm uses real templates for filtering, and to account for both phases of the matched filter response, both the real and imaginary parts of each template are filtered using Eq.~\ref{eq:mf}. For a template bank containing $M$ template waveforms, this results in $2M$ real template filters.

Modern template banks can contain millions of waveforms, with low-mass binary templates spanning hundreds of seconds. Direct computation of these cross-correlations for all templates is therefore computationally prohibitive, particularly in low-latency applications where results must be produced in near real time.

The LLOID algorithm addresses this challenge by dividing each template into time slices and compressing them with SVD, then processing each slice at the minimum sampling rate required to capture its frequency content \cite{cannon2012b}. By representing templates with a reduced set of orthogonal basis waveforms and employing a multirate filtering strategy, LLOID significantly reduces the computational cost while preserving sensitivity to gravitational-wave signals.

\subsubsection{SVD bank construction}\label{subsec:svd}
The SVD bank construction follows the LLOID algorithm, combining time slicing and SVD to compress the template bank. Templates are first divided into groups \cite{sakon, messick2017}. Within each group, templates are whitened and then divided into $S$ non-overlapping time slices, with time intervals $[t^0, t^1)$, $[t^1, t^2)$, \ldots, $[t^{S-1}, t^{S})$, where $t$ is measured as time before the merger and increases going backward in time, so that slice 0, spanning $[t^0, t^1)$, lies closest to the merger and is the highest-frequency slice \cite{cannon2012b}. The time slices within a group share the same time boundaries across templates. Each slice $s$ is downsampled to the Nyquist rate corresponding to its highest frequency content, giving the sequence of sample rates $f^0, f^1, \ldots, f^{S-1}$, where $f^0$ is the highest rate (slice 0) and the rates decrease for earlier, lower-frequency slices. This ensures each slice is sampled at the minimum rate needed for that slice.

For each time slice, the set of templates is arranged into a matrix, with real and imaginary components stacked, and then decomposed using SVD. This produces a set of orthonormal basis waveforms ranked by their significance. Only the leading bases are retained, as higher-order components contribute negligibly to the accuracy required for matched filtering:
\begin{equation}
    \hat{h}_i^s[k] \approx \sum_{l=0}^{L^s-1} v^s_{il} \sigma^s_{l} u^s_{l}[k],s\in[0, S-1],
\end{equation}
where $S$ is the number of time slices, with 0 representing the latest time slice. $u^s_{l}[k]$ are the orthonormal basis waveforms, $v^s_{il}$ are the reconstruction coefficients for template $i$, $\sigma^s_{l}$ are the singular values, ordered by their contribution to reconstructing $\hat{h}_i^s[t]$, and $L^s$ is the number of basis components in the time slice. The SVD procedure reduces the number of waveform filters by approximately two orders of magnitude in the GstLAL O4 template bank \cite{sakon}.

\subsubsection{Filtering workflow}
\begin{figure*}
    \centering
    \includegraphics[width=1\linewidth]{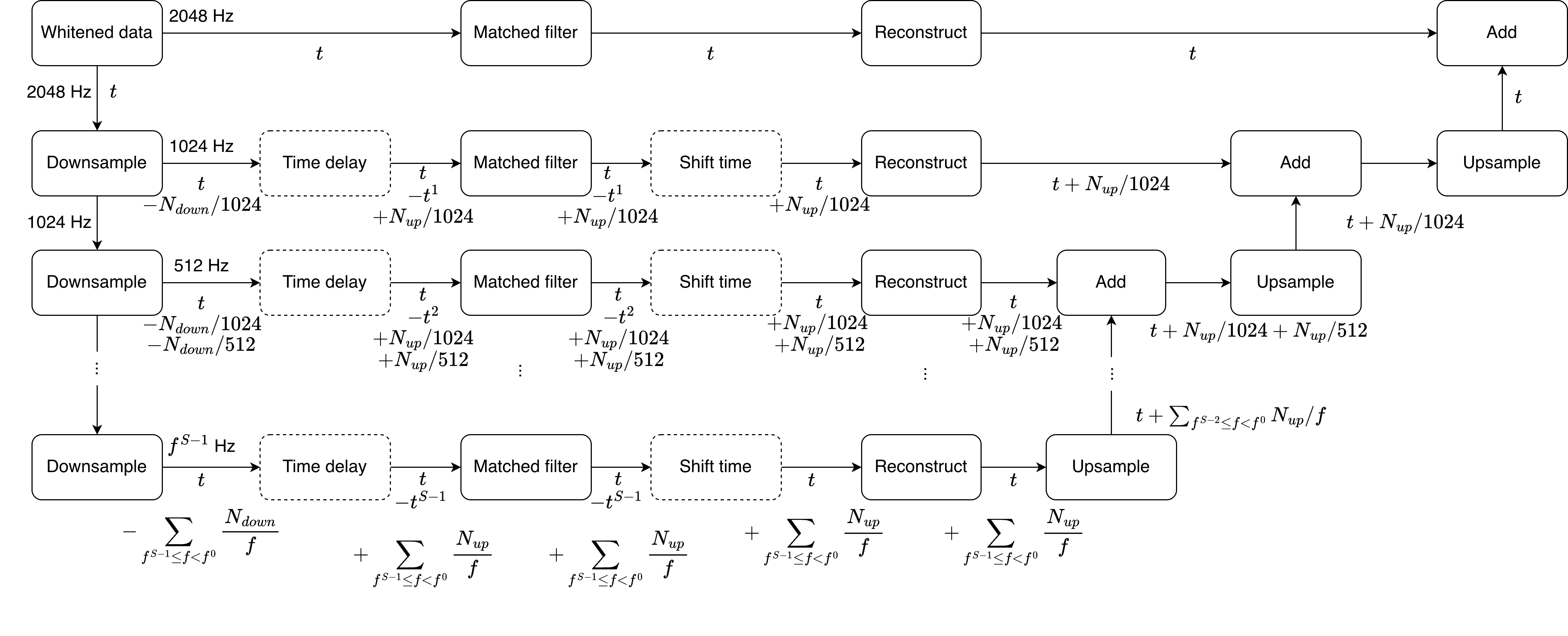}
    \caption{Diagram of the LLOID algorithm. The time evolution of the latest data sample at each processing stage is indicated alongside the transition arrows, where $t$ is the time of the latest input sample entering the LLOID algorithm. Whitened data are downsampled to match the sample rates of each time slices. After downsampling, the output lags the input by $N_{\mathrm{down}}/f$ seconds. Each time slice $s$ is then filtered using data delayed by the per-slice time delay $t^s$, where time slice $s$ spans $[t^{s}, t^{s+1})$, with $t$ measured as time before the merger and increasing going backward in time, plus an additional shift for the cumulative upsampling overlap between the slice's rate and the maximum rate. Following filtering, the output timestamp is advanced by the time delay associated with that slice. The filtered outputs are multiplied by the SVD reconstruction matrices to produce the physical SNR streams at each sample rate. The lowest-rate SNR stream is recursively upsampled and added to the higher-rate streams, where each upsampling stage introduces a latency of $N_{\mathrm{up}}/f$ seconds. This pre-synchronization of delays and time shifts ensures that the upsample-and-add process remains fully aligned, eliminating additional buffering and enabling the LLOID algorithm to operate with zero latency.}
    \label{fig:lloid}
\end{figure*}
\paragraph{Downsampling of whitened data}\label{sec:down}

Each template time slice requires a specific sample rate, so the input detector data is downsampled to match the rate of each time slice. This produces multiple downsampled streams that correspond to the different time slice rates.

Downsampling is implemented using a sinc-windowed sinc kernel:
\begin{equation}
g[k] =
\begin{cases} 
\dfrac{\sin\!\big(\pi (k-c)/f \big)}{\pi (k-c)/f} \cdot \dfrac{\sin\!\big(\pi (k-c)/c \big)}{\pi (k-c)/c}, & k \neq c, \\[0.5em]
1, & k = c,
\end{cases}\label{eq:sinc}
\end{equation}
where \(f\) is the downsampling factor (restricted to powers of two) and \(c\) is the half length of the filter. The total kernel length is $2c + 1$, and for downsampling $c=f N_{down}$,
where $N_{down}$ is the half filter length at the target rate, which is typically set to 32 samples in GstLAL, and determines the latency introduced by the downsampling filter.

\paragraph{Filtering and reconstruction}

After the input data for a given time slice $s$ is downsampled to match its sample rate, it is filtered by cross-correlating with the basis templates $u_l^s[n]$:
\begin{equation}
y_l^s[k] = \sum_{n=0}^{N^s-1} u_l^s[n] \, d^s[k+n],
\end{equation}
where $y_l^s[k]$ is the filtered output for the basis component $l$ in time slice $s$, and $N^s$ is the length of the basis template for that time slice.

Once all basis components are filtered, the original templates are reconstructed by combining these components using their singular values $\sigma_l^s$ and reconstruction coefficients $\nu_{il}^s$:
\begin{equation}
y{'}_i^s[k] = \sum_{l=0}^{L^s-1} \sigma_l^s \, \nu_{il}^s \, y_l^s[k].
\end{equation}

This reconstruction produces the SNR time series for all templates at the corresponding sample rate of the time slice.

\paragraph{Upsampling SNR segments}

After filtering and reconstructing each downsampled time slice, the outputs must be upsampled to the original sample rate. Standard upsampling typically inserts zeros between samples and convolves with a kernel, but SGNL (following GstLAL) uses a polyphase method to avoid multiplying by zeros.

The upsampling kernel is also a sinc-windowed sinc kernel, as defined in Eq.~\ref{eq:sinc}, with the total kernel length defined as $2 f N_{up} + 1$,
where $N_{up}$ is the half-length at the original sample rate for the upsampling kernel, and is set to 8 samples. To perform polyphase interpolation, the kernel is divided into $f$ sub-kernels:
\begin{equation}
z_j[k] = g[k f + j], \quad j = 0, \dots, f-1, \quad k = 0, \dots, 2c/f.
\end{equation}
Each sub-kernel is convolved directly with the input, and the outputs are interleaved to produce the final upsampled SNR time series:
\begin{equation}
y[i] =
\begin{cases}
x[i/f], & i \bmod f = 0,\\[6pt]
\displaystyle
\sum_{k=0}^{2c/f}
x[\lfloor i/f \rfloor - k]\, z_{\,i \bmod f}[k],
& \text{otherwise}.
\end{cases}
\label{eq:upsample}
\end{equation}
where $\lfloor \cdot \rfloor$ denotes floor division.

\paragraph{Combining SNR segments}

The SNR segments are combined recursively, starting from the lowest-rate time slice. Each segment is upsampled to the next higher sample rate and then added to the SNR contribution from the higher-rate slice. This process is repeated through all slices to construct the full SNR time series. Each time slice has a known time delay from the LLOID decomposition, and as long as these delays exceed the combined lengths of the downsampling and upsampling filters, no additional latency is introduced.

\subsubsection{Extensions in SGNL}
\paragraph{Pre-synchronization of time slices}\label{sec:sync}

In SGNL, time slices are pre-synchronized before filtering rather than only aligned after the LLOID algorithm output.  
For each time slice $s$, SGNL identifies exactly which segment of the downsampled input $x^s[k]$ is needed so that, after filtering, reconstruction, and upsampling, the slice’s contribution aligns precisely with the high-rate SNR output. This avoids trimming outputs or waiting for extra buffers.

Two main factors determine the segment:
\begin{enumerate}
    \item The time delay $t^s$ of time slice $s$, where time slice $s$ spans $[t^{s}, t^{s+1})$, with $t$ measured as time before the merger and increasing going backward in time.
    \item The cumulative half-lengths of all upsampling kernels between the slice’s rate $f^s$ and the maximum rate $f^0$.
\end{enumerate}
SGNL pre-synchronizes the slices by applying both offsets before filtering, identifying the exact portion of each downsampled stream to read so the slices align during the upsample-and-add without trimming or buffering.

The recursive reconstruction of the SNR contribution from slice $s$ can be expressed as:
\begin{align}
\rho^s[k] &= (H^{\uparrow}\rho^{s+1})[k] 
    + \sum_{l=0}^{L^s-1} \sigma_l^s \, \nu_{il}^s 
    \sum_{n=0}^{N^s-1} u_l^s[n] \notag\\
&\quad \times d^s \Bigg[ k + n - t^s
    + \sum_{\substack{f^s \le f < f^0}} \frac{N_{\text{up}}}{f} \Bigg],
\end{align}
where $H^{\uparrow}$ is the upsampling operator that maps a slice to the next-higher sample rate with Eq. \ref{eq:upsample}, and all other symbols are as defined in previous sections.

The final term, $\sum_{f^s \le f < f^0} N_{\text{up}}/f$, is the additional shift that accounts for the upsampling overlap between the slice's rate and the maximum rate. GstLAL applies only the slice delay $t^s$. The upsampled streams are then combined by the streaming framework, which buffers and queues each stream to align them by timestamp before summing. This buffering can introduce additional latency. SGNL instead accounts for the upsampling overlap explicitly when selecting which samples each slice reads during the matched filtering, synchronizing each slice in advance and avoiding the additional latency. Fig. \ref{fig:lloid} shows the LLOID algorithm as implemented in the SGNL pipeline.

\paragraph{Multidimensional filtering}

In SGNL, filtering can be performed on multiple SVD banks in parallel by stacking the bases and reconstruction matrices into tensor objects, which are multidimensional arrays. The retained bases and their reconstruction coefficients after SVD are grouped by sample rate. During initialization, bases from slices with the same rate are stacked into a tensor, as are their reconstruction coefficients. Each unique sample rate has a corresponding downsampled data stream, and the relevant time segments for each slice are aligned and stacked accordingly, as detailed in Sec. \ref{sec:sync}. This tensorized layout enables SGNL to perform filtering and matrix multiplication across multiple slices and SVD banks in parallel with PyTorch’s multibatch and multichannel operations, thereby optimizing GPU performance.

\subsection{Event identification}\label{sec:event}
\subsubsection{Trigger identification}
\begin{figure}
    \centering
    \includegraphics[width=1\linewidth]{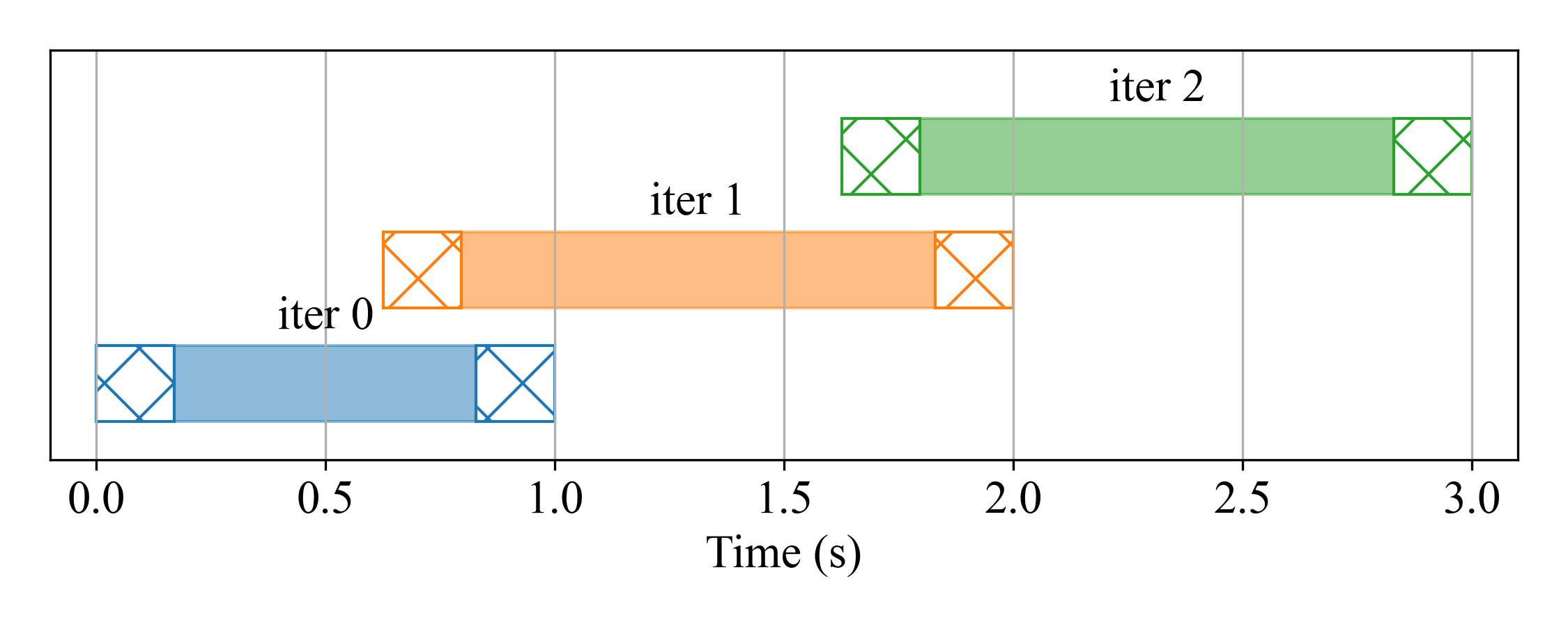}
    \caption{Overlapping and streaming behavior during the trigger and coincidence identification stage. Three consecutive iterations of the pipeline are shown, labeled above each block. Shaded regions mark the trigger-finding windows, and cross-hatched regions show the padding regions used for the $\xi^2$ calculation (350 samples in this example, corresponding to an autocorrelation length of 701 samples). The trigger-finding windows overlap by the sum of the maximum light-travel time between detectors (0.0273~s for an LIGO Hanford (H), LIGO Livingston (L), and Virgo (V) search) and a 0.005~s coincidence buffer. Each trigger finding window therefore spans one second plus the overlap interval. Input data are in one-second strides, thus the first window is shortened to produce output as soon as data become available, minimizing additional latency.}
    \label{fig:coinc}
\end{figure}
After the LLOID filtering algorithm, the $2M$ real template filters produce $2M$ SNR time series, with the real and imaginary components corresponding to the SNR outputs of each of the $M$ template waveforms.  For each detector and template, the trigger is defined as the maximum modulus of the complex SNR that exceeds a threshold of $\rho = 4$. The corresponding time, SNR, and phase are recorded for each trigger. 
This trigger identification process is performed within each one-second segment, including a small buffer to account for edge effects during coincidence formation (see Sec. \ref{sec:coinc} and Fig. \ref{fig:coinc}). This maximization is implemented using PyTorch operations and executed on the GPU when GPU mode is enabled.

\subsubsection{Signal-Based Consistency Test}

For each SNR peak identified during the trigger identification stage, a signal-consistency statistic is computed as
\begin{equation}
\xi^2_j = \frac{\int_{-\delta t}^{\delta t} dt\left| z_j(t) - z_j(0) R_j(t) \right|^2 }
{\int_{-\delta t}^{\delta t} dt\left( 2 - 2 \left| R_j(t) \right|^2 \right)},j\in[0,M-1],
\end{equation}
where $\xi^2_j$ quantifies the consistency of the SNR peak with the expected complex template autocorrelation. Here, $z_j(t)$ denotes the complex matched-filter SNR time series, $z_j(0)$ its peak value at the trigger time, and $R_j(t)$ the complex template autocorrelation function, normalized such that $R_j(0) = 1$ \cite{messick2017}. The integration half-width $\delta t$ corresponds to half the full autocorrelation length. The full autocorrelation length is 351 samples for high-mass templates and 701 samples for low-mass templates. This defines the segment of the SNR time series used in the $\xi^2_j$ computation and contributes to the overall pipeline latency. The streaming and overlap behavior are illustrated in Fig.~\ref{fig:coinc}, where the cross-hatched region represents the padding interval for the autocorrelation length calculation and indicates the associated processing latency.

To compute this statistic, the autocorrelation functions are stacked across SVD banks, and the corresponding segment of the SNR time series, centered on the trigger, is extracted with the same length as the autocorrelation across templates. The calculation is performed using tensor operations and executed on the GPU when GPU mode is enabled. Evaluated for each trigger, $\xi^2_j$ provides a waveform-consistency measure that complements SNR and detector coincidence in event ranking.

\subsubsection{Coincidence Formation} \label{sec:coinc}

To suppress false alarms, candidate events must occur in temporal coincidence across multiple detectors. The event time for each single-detector trigger is defined as the time of the SNR peak. For every trigger in one detector, the pipeline checks if triggers in the other detectors are within a coincidence window that accounts for the gravitational-wave travel time between sites. An additional buffer of 0.005~s is included to accommodate timing variations. To minimize edge effects, the trigger-finding windows overlap by the sum of the maximum light-travel time and the coincidence buffer. Fig.~\ref{fig:coinc} illustrates the small overlap between adjacent trigger-finding windows (shaded regions). The coincidence check is implemented using tensor operations and executes efficiently on the GPU when GPU mode is enabled.

\subsubsection{Clustering}\label{sec:clustering}
After coincidence formation, each event may consist of a variable number of coincident triggers across detectors. Within each subbank, events are then clustered, and only the event with the highest network SNR is retained, ensuring that at most one event is kept per subbank. In GPU mode, the selected events are transferred to the CPU for subsequent processing. The clustering uses the same time window as the trigger-finding stage, removing closely spaced redundant events.

\subsubsection{Significance estimation}

SGNL refactored GstLAL's likelihood ratio calculations for ranking statistics into a modular library, STRIKE \cite{strike}. This calculation requires collecting background noise distributions for each detector and for each SVD bin. Background collection uses the SNR and $\xi^2$ values of triggers that exceed the SNR threshold during the trigger-identification stage. We only include single-detector triggers that occurred while multiple detectors were operating.

In multi-SVD bank mode, the background is stored in a multi-dimensional tensor initialized at startup. Each SVD bank accumulates its background triggers independently. For each trigger-finding window, binning is performed in parallel across SVD banks. The resulting counts are continuously added to the persistent background tensor. These operations run on the GPU when GPU mode is enabled. The background is snapshotted every four hours, and is smoothened by the Gaussian kernel as kernel density estimation. At this point, the background tensor is transferred to the CPU.

For each event that survives the clustering stage (Sec. \ref{sec:clustering}), the likelihood ratio is calculated using the snapshotted background tensor of the specific bin that the event belongs to. The persistent background tensor on the GPU (in GPU mode) continues collecting new background data.

Each snapshot of the background is saved to disk and also posted via a Bottle \cite{bottle} route. A companion program queries this background data through Bottle services, draws samples from the SNR-$\xi^2$ distributions from each SVD bin across all inspiral programs, and constructs a marginalized background model across SVD bins. The resulting marginalized background is saved to disk. Inspiral jobs check for updated files and reload them in memory at every snapshot interval. FAR assignments for events with likelihood ratio evaluations are then calculated using the marginalized background.

Events and their associated triggers are saved to an in-memory SQLite database. At each snapshot interval, the triggers are also saved to disk.

\subsubsection{Event candidate alerting}\label{sec:alert}

After the likelihood ratio is calculated and FAR is assigned, events that pass the upload threshold are internally aggregated across subbanks to select a local candidate within a program. If the FAR is below the public alert threshold, the event with the maximum SNR is selected. If the FAR is above the threshold, the event with the minimum FAR is chosen. The local candidate is sent through Kafka. An auxiliary program further aggregates local candidates across all inspiral programs and sends the global candidate to GraceDB \cite{ewing2024}.

During internal aggregation, the local candidate may include triggers from only a subset of the observing detectors. For example, if H, L, and V are observing, the local candidate might initially contain triggers only from H and L. To provide complete SNR time-series information for downstream processing, we search for a corresponding trigger in the missing detector that lies within the light-travel-time window, and add it to the local candidate. The SNR time series for each trigger forming the local candidate are then extracted and stored in the local-candidate metadata before it is sent out. This ensures full SNR time series information is available for downstream skymap generation.

\subsubsection{Monitoring metrics}

Metrics from each event, including SNR, likelihood ratio, and FAR, and latency at various pipeline stages, are sent to Kafka topics. Metrics from multiple inspiral programs are aggregated and stored in an InfluxDB \cite{influxdb}. Grafana \cite{grafana} panels are then used to visualize these metrics over time and monitor the pipeline’s performance and stability.

\section{Mock data challenge results}\label{sec:results}
To assess the performance of the SGNL online analysis and compare it with GstLAL, we ran an analysis on an MDC. The MDC consists of 40 days of O3 H, L, and V data, from Jan 05 23:59:42 GMT 2020 to Feb 14 23:59:42 GMT 2020. The data are replayed with timestamps shifted to the present to simulate live streams. Alongside the strain data, additional channels containing simulated CBC waveforms for binary neutron stars (BNS), neutron star black hole binaries (NSBH), and binary black holes (BBH) were streamed in parallel. Injections were added roughly every minute, for a total of about 50,000 injections. A detailed description of the MDC and the injection set can be found in \cite{mdc}.

We ran the GstLAL and SGNL analyses in parallel on the same computing cluster and uploaded their low-latency events to the same GraceDB instance. The O4 template bank is split into two ``checkerboarded'' halves \cite{sakon}. This checkerboarding process takes every other adjacent template in the full bank, forming two banks that cover the same parameter space. GstLAL and SGNL processed the same checkerboard of the template bank, ensuring that they used the same input templates. GstLAL was operated in its end-of-O4 configuration. Both GstLAL and SGNL were deployed on CPUs.

In this section we aim to compare the performance between SGNL and GstLAL on the MDC. We will first analyze the event recovery of known gravitational-wave events identified in O3 within the duration of the MDC data. Next, we will analyze the injection recovery comparison in terms of the sensitive spacetime volume and injection parameters. Finally, we will highlight the difference in latency performance between the pipelines.

\subsection{Gravitational wave events}
\begin{figure}
    \centering
    \includegraphics[width=\linewidth]{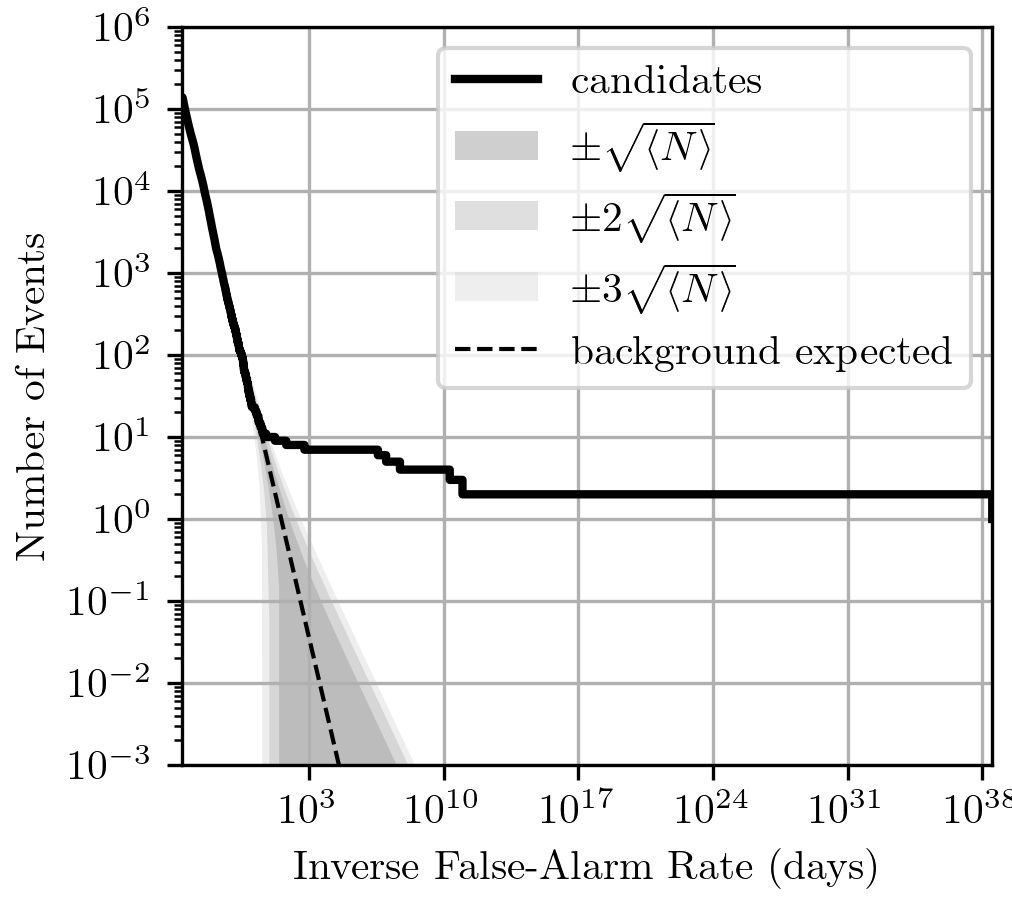}
    \caption{Number of events versus IFAR in days. Dashed line: expected background distribution. Shaded areas: 1, 2, and 3 $\sigma$ uncertainties. Black line: number of candidates observed.}
    \label{fig:money-plot}
\end{figure}

There are nine gravitational-wave events previously published in GWTC-3 \cite{gwtc-3} that are in the MDC data. Table \ref{table:events} summarizes the SNR and FAR results for all nine gravitational-wave events in the MDC for both GstLAL and SGNL. The preferred event was selected with the criteria of the maximum SNR event among the events with FAR below the public alert threshold, otherwise the minimum FAR event was selected. The number of events versus inverse false-alarm rate (IFAR) plots for the SGNL MDC are shown in Fig. \ref{fig:money-plot}.

For eight of the nine gravitational-wave events, SGNL recovered network SNRs within $<1\%$ of GstLAL's values and obtained consistent FARs. The remaining event, GW200202\_154313, showed a lower SNR and a FAR that was orders of magnitude higher in the SGNL MDC compared to GstLAL. Comparing the event recovered from SGNL and GstLAL, we found that they were recovered with different templates, although both templates belong to the same SVD subbank. Offline re-filtering showed that this difference comes from the trigger-finding window boundaries: the SNR maximization step selects different triggers when the boundaries shift, especially for triggers close to threshold. This accounts for the SNR difference between the SGNL event and the GstLAL event, and contributes to the FAR discrepancy as well. This is a symmetric effect. The window boundaries could equally disadvantage GstLAL for a different near-threshold event, so it does not systematically favor either pipeline.
Although this event is not recovered as significantly in SGNL as it is in GstLAL, injection studies demonstrate that the search sensitivity between SGNL and GstLAL is consistent within statistical and systematic uncertainties (see Sec.~\ref{sec:inj}). This indicates that the behavior of this particular event does not reflect a systematic issue.

\begin{table*}
\caption{Comparison of gravitational-wave event candidates previously reported in GWTC-3 within the MDC time span, between GstLAL and SGNL. The “found inst.” column lists the instruments that identified each event with a trigger SNR $\ge$ 4.0. All reported SNRs are network SNRs.\label{table:events}}
\centering
\begin{ruledtabular}
\begin{tabular}{lccccc}
& &   \multicolumn{2}{c}{GstLAL MDC} & \multicolumn{2}{c}{SGNL MDC}\\
\cline{3-4}\cline{5-6}
Name & Found Inst.     & SNR & FAR (yrs$^{-1}$)  & SNR & FAR (yrs$^{-1}$) \\
\hline
GW200112\_155838 & L1 &  18.46 &9.69$\times10^{-9}$ & 18.46 & 1.11$\times10^{-8}$ \\
GW200115\_042309 & H1L1 &  11.48 & 2.44$\times10^{-5}$ & 11.51 & 1.68$\times10^{-5}$\\
GW200128\_022011 & H1L1 &   9.98 & 3.78$\times10^{-5}$&9.98& 3.52$\times10^{-5}$ \\
GW200129\_065458 & H1L1V1 & 26.30 &9.43$\times10^{-40}$ &26.27& 1.10$\times10^{-36}$\\
GW200202\_154313 & H1L1 &10.57 & 2.10$\times10^{-2}$  &10.13& 3.52$\times10^{2}$ \\
GW200208\_130117 & H1L1 & 10.54 &3.44$\times10^{-3}$ & 10.53& 1.91$\times10^{-4}$  \\
GW200208\_222617 & H1L1 & 8.03 & 5.16$\times10^{2}$ &	8.03 & 8.20$\times10^{2}$ \\
GW200209\_085452 & H1L1 & 9.96 & 3.45$\times10^{-1}$ &9.96 & 8.95$\times10^{-1}$ \\
GW200210\_092254 & H1L1 & 9.26 & 4.33$\times10^{2}$&9.21 & 6.63$\times10^{2}$ \\
\end{tabular}
\end{ruledtabular}
\end{table*}

\subsection{Retraction}
In addition to the known gravitational-wave events, we identify ``retraction-level candidates,'' defined as events with a FAR below one per year for the MDC~\cite{ewing2024}. Both GstLAL and SGNL found one such event, corresponding to the same GPS time in the original O3 data. GstLAL recovered this event in the L1 detector with an SNR of 15.29, and SGNL found it in the same detector with an SNR of 15.30. The FAR values are $1.12\times10^{-8}$ and $1.73\times10^{-8}$~per~year for GstLAL and SGNL, respectively, both below the one-per-year threshold. This event also appeared in previous GstLAL MDC studies~\cite{ewing2024}. As shown in Fig.~3 of~\cite{ewing2024}, the corresponding spectrogram exhibits scattering glitches, and because the event was recovered in only one detector while all three were operational, it is likely of terrestrial origin. Since the purpose of this MDC comparison is to assess consistency between GstLAL and SGNL, the recovery of this retraction event by SGNL is expected.

\subsection{Recovered injections}\label{sec:inj}
\subsubsection{Search Sensitivity}

\begin{figure}
    \centering
    \includegraphics[width=\linewidth]{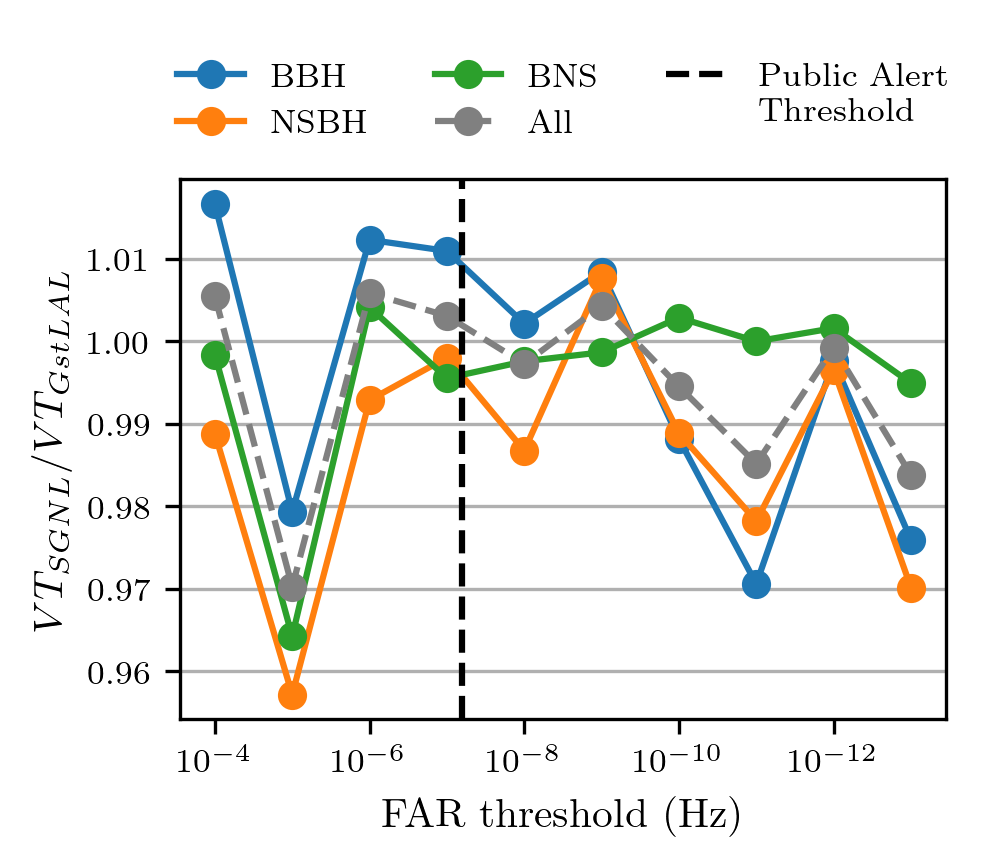}
    \caption{$\langle VT\rangle$ ratio between SGNL and GstLAL at the end of the MDC for each source class. $VT_{SGNL}$ is the $\langle VT\rangle$ for the SGNL analysis, while $ VT_{GstLAL}$ is the $\langle VT\rangle$ for the GstLAL analysis. BBH $\langle VT\rangle$ is shown in blue, NSBH $\langle VT\rangle$ in orange, BNS $\langle VT\rangle$ in green, and the combined $\langle VT\rangle$ across all sources is shown in the grey dashed line.
    The vertical dashed line represents the public-alert FAR threshold, at which low-latency public alerts are sent out. The $\langle VT\rangle$ ratio between SGNL and GstLAL is within $\sim1\%$ at the public-alert threshold.}
    \label{fig:vt-ratio}
\end{figure}

We use the spacetime volume, or sensitivity volume-time (VT), to quantify the analysis sensitivity. 
The comoving volume is defined as \cite{Hogg:1999ad, Chen:2017wpg}
\begin{equation}
    V_c = 4 \pi D_H \int_0^z dz \frac{(1+z) D_A^2}{E(z)},
\end{equation}
where $z$ is the redshift, $D_H$ is the Hubble distance, $D_A$ is the angular distance, and $E(z)$ is the Hubble parameter. 
Multiplying the comoving volume by the total observing time gives the searched spacetime volume. 
The injection VT, $\langle VT \rangle_{\rm inj}$, is calculated using the maximum redshift of the injections and the time span they cover. 
The analysis VT, $\langle VT \rangle$, is defined as the fraction of recovered injections multiplied by the injection VT \cite{ewing2024}:
\begin{equation}
    \langle VT \rangle = \frac{\text{found}}{\text{total}} \langle VT \rangle_{\rm inj}.
\end{equation}
The VT ratio between analyses is defined as the ratio of the number of recovered injections.

Fig. \ref{fig:vt-ratio} shows the VT ratio between the SGNL analysis and the GstLAL analysis at different FAR thresholds, which are calculated as the ratio of number of found injections, with the ``found" criteria being the injections that are below the given FAR threshold. The VT ratios are calculated for difference source classes, BNS, NSBH, and BBH.
The vertical dashed line denotes the public-alert FAR threshold, at which public alerts are distributed, and the VT ratio is within $\sim1\%$.
Contributions to the discrepancies are the lack of the sub-sample interpolation in SNR, the difference in analysis operating time, and the occurrence of dropped data. We conclude that the $\langle VT\rangle$ of SGNL is consistent with GstLAL within statistical and systematic uncertainties.

\subsection{Upload latency}
\begin{figure}
    \centering
    \includegraphics[width=\linewidth]{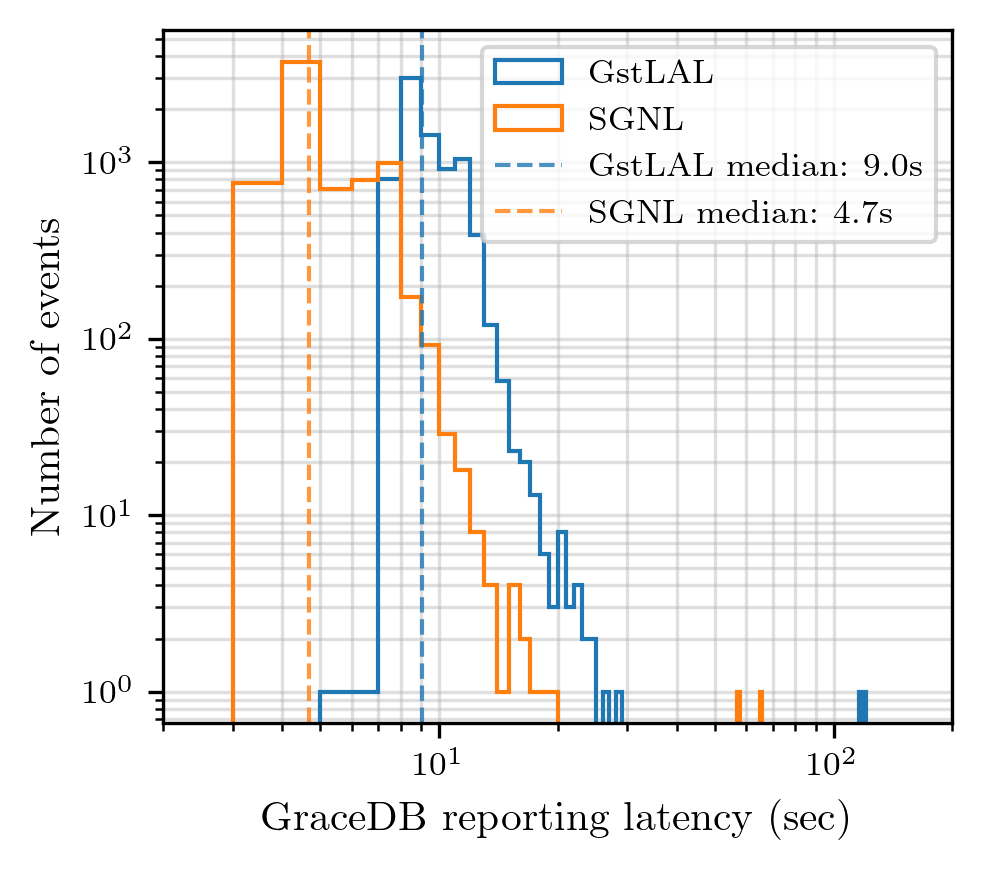}
    \caption{Histograms of GraceDB reporting latencies. The blue histogram shows latencies for the GstLAL analysis, while the orange histogram shows latencies for the SGNL analysis. Dashed lines indicate the median latency for each pipeline: blue for GstLAL and orange for SGNL.}
    \label{fig:latency}
\end{figure}

Fig.~\ref{fig:latency} shows the histogram of GraceDB reporting latencies for GstLAL and SGNL. The GraceDB reporting latency is defined as the time when GraceDB receives an event minus the event coalescence time.
GstLAL's latency histogram shows a peak around 8--9 seconds, with the lowest latency at 6.9 seconds.
There is a small bump around 11--12 seconds,
which is caused by the downstream event aggregation process, where the aggregator waits two seconds to combine events \cite{ewing2024}.
SGNL's latencies peak around 4--5 seconds, with the shortest latency at 3.4 seconds. There is also a small bump around 7--8 seconds, caused by the same downstream aggregation process.
For GstLAL, the median latency is 9.0 seconds, while SGNL has a median latency of 4.7 seconds, representing a 48\% reduction.

\section{Conclusion} \label{sec:conclusion}

We have presented SGNL, a scalable, low-latency gravitational-wave detection pipeline for compact binary mergers. SGNL modernizes the GstLAL framework by leveraging a Python-based streaming architecture, SGN, enabling modular, flexible, and GPU-accelerated processing while preserving the physics-proven methods of matched filtering and likelihood-ratio ranking. Key innovations include pre-synchronization of LLOID time slices, multidimensional tensorized filtering, and optimized data streaming, all of which contribute to low-latency operation without sacrificing accuracy.

Performance evaluation using a MDC demonstrates that SGNL reliably recovers known gravitational-wave events and simulated injections, with sensitivities comparable to GstLAL, within expected uncertainties using a single checkerboard. Notably, SGNL reduced GstLAL’s median latency from 9.0 seconds to 4.7 seconds, which is a 48\% reduction.

Looking ahead, SGNL offers a flexible platform for future enhancements, including expanded parameter-space coverage and integration with machine learning techniques. Its design ensures that low-latency detection and multimessenger alerts can be performed efficiently, supporting rapid follow-up and maximizing the scientific output of the global gravitational-wave network.

\begin{acknowledgments}
This material is based upon work supported by NSF's LIGO Laboratory which is a major facility fully funded by the National Science Foundation.
This research has made use of data, software and/or web tools obtained from the
Gravitational Wave Open Science Center (https://www.gw-openscience.org/ ), a
service of LIGO Laboratory, the LIGO Scientific Collaboration and the Virgo
Collaboration.  We especially made heavy use of the LVK Algorithm
Library~\cite{lalsuite}. LIGO Laboratory and Advanced LIGO are funded
by the United States National Science Foundation (NSF) as well as the Science
and Technology Facilities Council (STFC) of the United Kingdom, the
Max-Planck-Society (MPS), and the State of Niedersachsen/Germany for support of
the construction of Advanced LIGO and construction and operation of the GEO600
detector. Additional support for Advanced LIGO was provided by the Australian
Research Council.  Virgo is funded, through the European Gravitational
Observatory (EGO), by the French Centre National de Recherche Scientifique
(CNRS), the Italian Istituto Nazionale di Fisica Nucleare (INFN) and the Dutch
Nikhef, with contributions by institutions from Belgium, Germany, Greece,
Hungary, Ireland, Japan, Monaco, Poland, Portugal, Spain.

The authors are grateful for computational resources provided by the 
the Pennsylvania State University's Institute for Computational and Data
Sciences gravitational-wave cluster, and the LIGO Lab cluster at the LIGO Laboratory, and supported by 
the National Science Foundation awards 
OAC-2103662, PHY-2308881, PHY-2011865, OAC-2201445, OAC-2018299, PHY-0757058, PHY-0823459, PHY-2207728, PHY-2513124, PHY-2110594, and PHY-2513358.
CH Acknowledges generous support from the Eberly College of Science, the 
Department of Physics, the Institute for Gravitation and the Cosmos, and the 
Institute for Computational and Data Sciences.

\end{acknowledgments}
\bibliography{references.bib}

\end{document}